\documentclass[prc,twocolumn, showpacs,superscriptaddress,preprintnumbers,amsmath,amssymb,nofootinbib]{revtex4}

\usepackage[dvipdfmx]{graphicx}
\usepackage{dcolumn}
\usepackage{bm}
\usepackage{amsmath}

\usepackage[normalem]{ulem}  
\usepackage[dvips]{color} 

\newcommand{\changed}[1]{{#1}}

\renewcommand\sout{\bgroup \color{red} \ULdepth=-.5ex \ULset}

\begin{document}


\title{

Estimation of electric conductivity of the quark gluon plasma \\
via asymmetric heavy-ion collisions

}

\author{Yuji~Hirono}
 \email{yuji.hirono@stonybrook.edu}
\affiliation{
Department of Physics and Astronomy, Stony Brook University, 
Stony Brook, New York 11794-3800, USA
}

\author{Masaru~Hongo}
 \email{hongo@nt.phys.s.u-tokyo.ac.jp}
\affiliation{
Department of Physics, The University of Tokyo, 
Hongo~7-3-1, Bunkyo-ku, Tokyo 113-0033, Japan
}
\affiliation{
Theoretical Research Division, Nishina Center, RIKEN, Wako 351-0198,
 Japan
}
\affiliation{
Department of Physics, Sophia University, Tokyo 102-8554, Japan
}

\author{Tetsufumi~Hirano}
\email{hirano@sophia.ac.jp}
\affiliation{
Department of Physics, Sophia University, Tokyo 102-8554, Japan
}

\date{\today}

\begin{abstract}
We show that in asymmetric heavy-ion collisions, especially off-central Cu+Au
collisions, a sizable strength of electric field directed from Au
 nucleus to Cu nucleus is
generated in the overlapping region, because of the difference in the
number of electric charges between the two nuclei.
This electric field would induce an electric current in the matter
 created after the collision, which results in a dipole deformation
 of the charge distribution.
The directed flow parameters $v_1^{\pm}$ of charged particles 
turn out to be sensitive to the charge dipole and provide us with
 information about electric conductivity of the quark gluon plasma.
\end{abstract}

\maketitle

\emph{Introduction.}---
The quark gluon plasma (QGP), which consists of deconfined quarks and gluons, 
 is expected to have filled the early universe \cite{Yagi:2005yb}.
We are now at the stage to study properties of the QGP
experimentally 
through the relativistic heavy-ion collisions 
using Relativistic Heavy Ion Collider (RHIC) at BNL
and the Large Hadron Collider (LHC) at CERN.
One of the most interesting observations is the very strong elliptic
flow in off-central collisions, which indicates a very small 
ratio of shear viscosity to entropy density, $\eta / s$
\cite{Adams:2005dq,Adcox:2004mh,Romatschke:2007mq,Song:2010mg}. 
We are now trying to learn more detailed properties of the QGP by
constraining transport coefficients such as shear viscosity, bulk
viscosity, and charge diffusion constants.

In this Letter, 
we propose a new way of estimating the electric conductivity $\sigma$ of
the QGP
via asymmetric nucleus-nucleus collisions at ultrarelativistic energies.
Theoretically, lattice QCD
simulations \cite{Gupta:2003zh, Aarts:2007wj, Ding:2010ga,Buividovich:2010tn}
and perturbative QCD calculations \cite{Arnold:2003zc}
have been utilized to estimate electric conductivity of the QGP.
So far, the estimated values of $\sigma$  have differed
significantly from each other
and experimental information is intently awaited. 
Very recently, asymmetric collisions between copper
(Cu) and gold (Au) nuclei have been performed
at RHIC, and the PHENIX Collaboration reported their first results
\cite{Huang:2012sc}.
We show that Cu+Au collisions can be useful
for extracting the electric conductivity of the QGP.
In off-central Cu+Au collisions, a substantial magnitude of
electric field directed from a colliding Au nucleus to Cu nucleus
is generated in the overlapping region.
This happens only when the two colliding nuclei carry different numbers
of electric charge 
\footnote{
The possibility of estimating the electric conductivity of the QGP is
suggested in Ref.~\cite{Bzdak:2011yy}. 
}.

This electric field would induce a current in the matter created after
the collision, resulting in a dipole deformation of the charge
distribution in the medium.
Later, the time evolution of the system is dominated by a 
strong radial flow, which is an outward collective motion of the medium.
Henceforth the charge asymmetry formed in the early stage is frozen.
Thus, we argue that charge-dependent directed flow
of the observed hadrons is sensitive to the charge dipole 
formed at the early stage, which
reflects the electric conductivity of the QGP.

Conventionally the electric conductivity of the QGP is 
estimated from experiments via the Kubo formula \cite{Kubo:1957mj}.
The production rate of thermal 
dileptons is expressed by the electric current-current correlation function 
\cite{McLerran:1984ay, Gale:1987ki}
and its small frequency region
is governed by the transport peak \cite{Forster:1990}.
Thus one can estimate the electric conductivity
through comparison of theoretical results with dilepton invariant mass spectra \cite{Akamatsu:2011nr}.
Compared to this method, the present approach is a rather direct one, 
in which the response of the matter to an applied electric field is
directly quantified.

The effects of transient strong electromagnetic fields havve been 
under intensive discussions recently, especially in the context of the
chiral magnetic effect \cite{Fukushima:2008xe,Burnier:2011bf,Burnier:2012ae}. 
So far, there has been no experimental evidence that
strong fields actually exist. Observation of a charge-dependent directed
flow would also provide evidence that a strong electromagnetic field
is actually created in heavy-ion collisions.

\emph{Electric fields in Cu+Au collisions.}---
Here, we show that, in off-central collisions between copper and gold nuclei,
a sizable strength of electric field is generated in the overlapping
region of two nuclei.
Because of the difference in the number of protons between the two nuclei,
the generated electric field tends to the copper nucleus.
The situation is different from the electromagnetic fields in the
collisions of the same species of nuclei \cite{Bzdak:2011yy,
Deng:2012pc}.
In symmetric collisions such as Au+Au or Cu+Cu, 
the event-averaged electric field does not
have a specific direction,
although the magnitude of the electric fields 
generated in each event
is
considerably large [$ |e \vec E| \sim |e \vec B| \sim O (m_\pi^2 ) $].
We have performed event-by-event
calculations of the electromagnetic fields in Cu+Au collisions
to show that there should be a significantly large copper-directed
electric field.

\begin{figure}[tbp]
 \begin{center}
  \includegraphics[width=65mm]{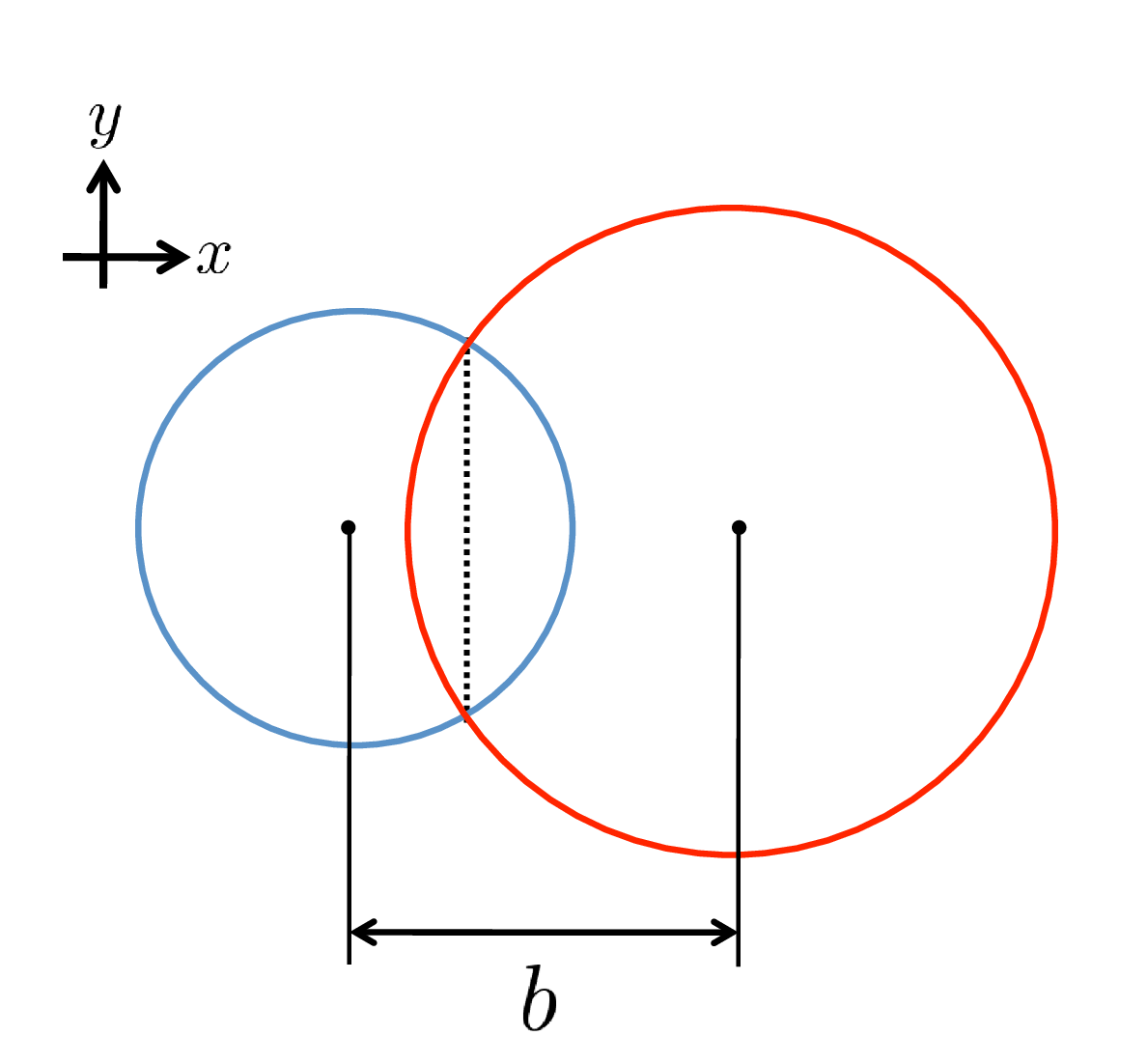}
 \end{center}
 \caption{
(Color online)
Transverse plane of off-central Cu+Au collisions with impact parameter
 $b$. 
The left (right) circle indicates the edge of the Cu (Au) nucleus.
}
\label{fig:coordinate}
\end{figure}

The electromagnetic fields are generated by the protons in nuclei.
If we regard protons as point particles, 
the electric and magnetic fields at a position $\vec x$ and time $t$
are written by the Li\'{e}nard-Wiechert potentials,
\begin{eqnarray}
 |e| \vec E (t, \vec x) &=& \alpha_{\rm EM} \sum_n \frac{1 - v_n^2 }
{R_n^3 \left[ 1 - (\vec R_n \times \vec v_n)^2 / R_n^2 \right]^{3/2}  }
\vec R_n, \\
 |e| \vec B (t, \vec x) &=& \alpha_{\rm EM} \sum_n \frac{1 - v_n^2 }
{R_n^3 \left[ 1 - (\vec R_n \times \vec v_n)^2 / R_n^2 \right]^{3/2}  }
 \vec v_n \times \vec R_n
, \nonumber \\
\end{eqnarray}
where 
$\vec R_n \equiv \vec x - \vec x_n(t)$ with $\vec x_n(t)$ the position
vector of the $n$-th proton, 
$\vec v_n$ is the velocity vector of the $n$-th proton, 
$|e|$ is the electric charge of a proton,
and 
$\alpha_{\rm EM}$ is the fine structure constant.
We define the origin of the spatial coordinate as the middle of the
 centers of the nuclei and $x$ and $y$ axes as in Fig.~\ref{fig:coordinate}
\footnote{
One may wonder whether we can take the origin of the azimuthal angle 
on the Au nucleus side in experiments.
According to a report from the PHENIX group \cite{Huang:2012sc},  
it is indeed possible to experimentally determine on which side the Au or Cu
nucleus is.
By measuring the spectators,
the origin of the azimuthal angle in the event plane of $v_1$ is
determined and is always taken on the Au-going side. 
}.
The summation is taken over all the protons in the colliding two nuclei.
The positions of the protons inside a nucleus 
are sampled from the
Woods-Saxon distribution with the standard parameters \cite{Alver:2008aq}.

\begin{figure}[tbp]
 \begin{center}
  \includegraphics[width=100mm]{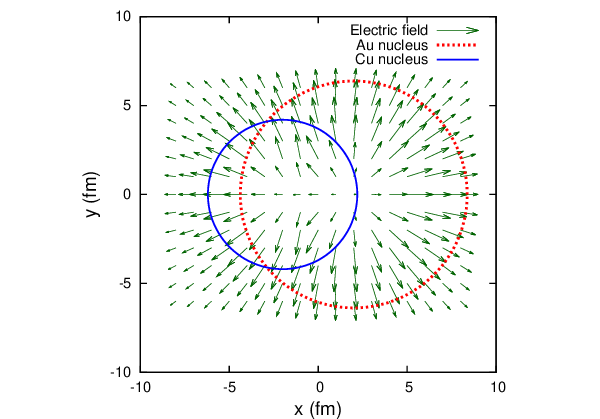}
 \end{center}
 \caption{
(Color online)
Event-averaged electric field in the transverse plane in off-central
Cu+Au collisions at $t=0$ (the collision time) with impact parameter $b= 4$ fm
at
$\sqrt{s_{NN}} = 200$ GeV.
Vectors are shown only in $|y|<6$ fm.
The average is taken over $10^4$ events.
}
\label{fig:Eaucu}
\end{figure}

Figure \ref{fig:Eaucu} shows 
the event-averaged electric fields in Cu+Au collisions
at impact parameter $b=4$ fm.
Each vector represents direction and magnitude
of the electric field at that point.
We find that the electric field in the central region
of the overlapping area 
has a specific tendency to go from Au to Cu.
Although the direction of electric fields fluctuates on an
event-by-event basis because of the fluctuation in the proton positions
inside colliding nuclei, 
the direction is correlated with the reaction plane for asymmetric
collisions.
The magnitude of the electric fields is as large [$|e \vec E| \sim
O (m_\pi^2 ) $] as the electric and magnetic fields in Au+Au collisions
at the same collision energy.

\begin{figure}[tbp]
 \begin{center}
  \includegraphics[width=80mm]{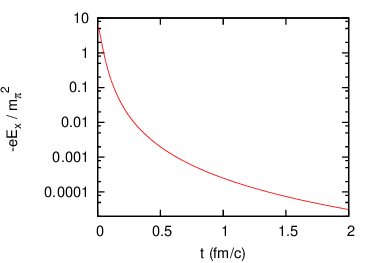}
 \end{center}
 \caption{(Color online)
Event average of the time evolution of electric fields in off-central Cu+Au
 collisions ($b=4$ fm) at $\vec x =\vec 0$. 
The average is taken over $10^4$ events.
}
\label{fig:E-time}
\end{figure}

We have also calculated the time dependence of the averaged electric
fields as shown in Fig.~\ref{fig:E-time}.
The strength of the fields decays  as the
spectators fly away. 
Nevertheless, it is notable
  that even at $t = 1 \ {\rm fm}/c$ the electric field is considerably
larger than the so-called ``critical field'' for electrons,
$|e|B_c=|e|E_c=m^2_e$ \cite{Schwinger:1951nm}.


\emph{Electric dipole of the plasma and charge-dependent directed flow.---}
The strong electric field toward the Cu nucleus at the early stage 
would induce an electric current in the medium that consists of the QGP 
after the thermalization time
\footnote{
The matter would be in the state of glasma 
before thermalized quark gluon plasma is formed. 
It is possible that the measured charge-dependent directed flow also
reflects the conducting property of such matter.
However, we expect that the conductivity of glasma is far smaller compared
to that of quark gluon plasma. 
That is because glasma basically does not have charged particles as a
constituent, while the QGP does.
The quarks and antiquarks in the QGP are deconfined, which makes QGP 
have a high conductivity.
That is why we assume that the charge asymmetry created in the
evolution originates mostly from the property of QGP.
}.
As a result, the charge distribution would be modified and 
a charge dipole would be formed.
One can expect that the dipole-like deformation of the charge
distribution at the early stage would also be present in the observed charge
distribution.
This is because the electromagnetic charge is an exactly conserved
quantity and an inhomogeneity relaxation of a conserved charge density takes a long time.
Once a radial flow starts, the medium expands rapidly and
the charge dipole created at the early stage would be frozen.
Thus,  we can reasonably assume that the dipole-deformation in the plasma remains intact
in the observed charge distribution.

The azimuthal angle distribution of the net charge, to the leading order
in the multipole expansion, 
can be written as 
\begin{equation}
\frac{ d \left(N_+ - N_-\right) }{d \phi} (\phi)
= \left( \bar N_+ - \bar N_-  \right)
\left( 1 + 2 d_e \cos \phi \right),
\label{eq:net-charge-v1}
\end{equation}
where 
the  azimuthal angle $\phi$ is measured from the $x$-axis and 
$\bar N_\pm$ is defined as the angle average of the number distribution,
\begin{equation}
 \bar N_\pm \equiv \int \frac{d\phi}{2\pi}
\frac{dN_\pm}{ d \phi}.
\end{equation}
The dipole deformation of the medium is quantified by the value of
$d_e$.
We assume that 
the azimuthal distribution of the total number of
particles is still written by $v_1$ without the effect of the
electromagnetic fields as 
\begin{equation}
\frac{ d \left(N_+ + N_-\right) }{d \phi} 
= \left( \bar N_+ + \bar N_-  \right)
\left( 1 + 2 v_1 \cos \phi  \right), 
\label{eq:total-v1}
\end{equation}
since electromagnetic fields are not expected to change the bulk flow
significantly.
From Eqs.~(\ref{eq:net-charge-v1}) and (\ref{eq:total-v1}), 
the distribution of charged particles can be written as 
\begin{equation}
 \begin{split}
 \frac{d N_+}{d \phi} &= 
\bar N_+ 
\left[
1+
 \frac{\bar N_+ + \bar N_- }{2 \bar N_+}
2 \left( v_1 + A d_e  \right) \cos \phi
\right]\\
&=
\bar N_+ 
\left\{
1+ 
2\left[ v_1 + A (d_e -  v_1)  \right] \cos  \phi
 + O\left[(A d_e)^2\right]
\right\},
\end{split}
\end{equation}
where we have defined the charge asymmetry parameter $A \equiv (\bar N_+ - \bar N_-)/(\bar N_+ + \bar N_-)$.
Similarly, 
\begin{equation}
\frac{d N_-}{d \phi} 
=
\bar N_-
\left\{
1+ 
2 \left[  v_1 - A (d_e -  v_1)  \right] \cos \phi
 + O\left[(A d_e)^2\right]
\right\}.
\end{equation}
Thus, the directed-flow coefficients $v_1$ for positively and negatively
charged particles are written as 
\begin{equation}
v_1^\pm =   v_1 \pm A d'_e,
\end{equation}
where we have defined $ d'_e \equiv d_e -  v_1$.
The values $v_1^\pm$ are linear functions of $A$ and their slopes are given by the
dipole-like deformation parameter $d'_e$, which is written as 
\begin{equation}
 d'_e = \frac{1} {\bar N_+ - \bar N_-} \int r dr d \phi \ 
 \cos \phi \ 
\left[
j_e^0(r,\phi) - j_{e, \vec E=\vec B=0}^0(r,\phi) 
\right],
\end{equation}
where $j_e^0(r,\phi)$($j_{e, \vec E=\vec B=0}^0(r,\phi)$) is the
transverse charge density in the presence (absence) of electromagnetic fields.

\emph{
Estimate of the charge-dependent directed flow.---
}
Let us make an order-of-magnitude estimate of the value of the
charge-dependent directed flow parameter $A d'_e$.
For that purpose, 
we first roughly evaluate the total charge that is transfered from
the gold-side to copper-side in the presence of an electric field.
The total charge $Q$ transfered across a plane $S$ from $t=0$ to $\tau$
is written as 
\begin{equation}
   Q = \int_0^\tau  d t \int_S  \vec J \cdot d \vec S
=\int_0^\tau  d t \int_S  \sigma  \vec E \cdot d \vec S,
\end{equation}
where we have used the constitutive relation $\vec J = \sigma \vec E$
with $\sigma$ the electric conductivity.
Let $S$ be the plane which includes the origin and is
perpendicular to the line which connects the centers of the two
colliding nuclei at $t=0$, the moment two nuclei contact.
Neglecting the space-time dependence of $\sigma$, $Q$ is rewritten as 
\begin{equation}
  Q \sim \sigma \tau \int_S \vec E \cdot d \vec S.
\label{eq:moved-charge-2}
\end{equation}
The integral in Eq.~(\ref{eq:moved-charge-2}) is
just the total electric flux that goes through the plane $S$. Hence, the
total transfered charge $Q$ is roughly
given by 
\begin{equation}
 \int_S \vec E \cdot d \vec S  \sim \frac{Z_{\rm Au} - Z_{\rm Cu}}{2}
  \frac{|e|}{\epsilon},
\end{equation}
where $Z_{\rm Au}$ and $Z_{\rm Cu}$ are the numbers of protons in the two
nuclei, and $\epsilon$
is the dielectric constant of the QGP.

According to lattice QCD simulations,
the electric conductivity of the QGP is estimated as 
\begin{equation}
 \sigma  \sim B \  C_{\rm EM} T, \quad C_{\rm EM} \equiv \sum_f e^2_f,
\end{equation}
where the sum in the electromagnetic vertex factor is taken over the
flavors and $B$ is a coefficient. 
If we consider $u$, $d$, and $s$ quarks,
$
  C_{\rm EM} = 8 \pi \alpha_{\rm EM} /3.
$
The value of the coefficient $B$ differs among calculations:
 $B \simeq 0.4$ in Refs.~\cite{Aarts:2007wj,Ding:2010ga} and  $B\simeq 7$ in Ref.~\cite{Gupta:2003zh}.
On the other hand, perturbative QCD calculations predicts  $\sigma
\simeq 6 T/e^2$ \cite{Arnold:2003zc}, which is much larger than the
values from lattice QCD simulations.

As for $\tau$, we take the time scale that the radial flow starts, $\tau \sim 1$ fm/$c$. 
If we take typical values for the other parameters, $T \sim 200 \ {\rm
MeV}$ and $\epsilon \sim 1$, 
the total transfered charge is estimated as 
\begin{equation}
 \begin{split}
  Q 
  &\sim B C_{\rm EM} T \tau \ \frac{Z_{\rm Au} - Z_{\rm Cu}}{2} \frac{|e|}{\epsilon} \\
&\sim B \cdot \frac{8 \pi}{3} \alpha_{\rm EM } \times 200 \ {\rm MeV} \times 1 \ {\rm
  fm}/c \times 25 |e| \\
& \sim 1.7 \ |e| \times B.
 \end{split}
\label{eq:total-charge}
\end{equation}

Now we can roughly evaluate $A d'_e$.
Let us choose the events in which the numbers of positive and negative
hadrons are equal, $\bar N_+ = \bar N_-$,  and 
assume that $n$ charges have been transfered by the electric field.
Then, the number $n$ can be written as  
\footnote{
Note that $A d'_e = A d_e$ for $\bar N_+=\bar N_-$.
}
\begin{equation}
\begin{split}
 n &=-\frac{1}{2} \int_{-\pi/2}^{\pi/2} d \phi \ 
\frac{d(N_+ - N_-) }{d \phi }
\\
&=
- 2  A d'_e \left( \bar N_+ + \bar N_-  \right).
\end{split}
\end{equation}
Therefore, the directed flow parameter $A d'_e$ is written by $n$ as 
\begin{equation}
A d'_e = - \frac{\pi n}{ N_{\rm tot} },
\label{eq:v1}
\end{equation}
where $N_{\rm tot} \equiv 2 \pi \left( \bar N_+ + \bar N_-  \right) $ is the total number of charged particles.
The number $n$ is related to the total transfered charge roughly as $n
\sim Q / |e|$.
Therefore, $A d'_e$, the charge-dependent part of the directed flow
parameter, and the electric conductivity of the plasma are parametrically
related as 
\begin{equation}
A d'_e \sim 
- \frac{\pi \sigma \tau}{N_{\rm tot}|e|} \int_S \vec E \cdot d \vec S.
\label{eq-charge-dep-v1}
\end{equation}
If one takes $N_{\rm tot} \sim 10^3$ and $n \sim 1$ [Eq.~(\ref{eq:total-charge})], 
the order of magnitude of the directed-flow parameter is estimated as
\begin{equation}
 A d'_e \sim - B \times 10^{-3} . 
\label{eq:dipole-estimate}
\end{equation}
This value would be within experimental reach if the parameter
$B$ is larger than of order unity.
Note that the value (\ref{eq:dipole-estimate}) is negative
since the electric field tends toward the Cu nucleus.
Although the estimate above is a crude one, 
 we can distinguish at least whether the created matter is in the
perturbative or non-perturbative regime by looking at the order of
magnitude of deference between $v^\pm_1$.
This is because perturbative calculations indicate significantly larger values of
$B(\sim 10^2)$ compared to lattice calculations ($B \sim 1 $).
This would indicate much progress 
compared to the current situation where little is known about the actual
conductivity of the matter created in heavy-ion collisions.

\begin{figure}[tbp]
 \begin{center}
  \includegraphics[width=70mm]{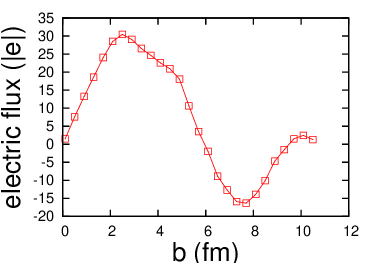}
 \end{center}
 \caption{
(Color online)
Event-averaged electric flux that goes through the thickest part of the almond-like
 shape of two overlapping nuclei, as a function of the impact parameter
 $b$. 
The value of the flux is the average over 2000 events for each $b$. 
}
\label{fig:flux}
\end{figure}
\changed{
As seen in Eq.~(\ref{eq-charge-dep-v1}) , the magnitude of the electric flux which goes 
through the QGP is an important quantity. We calculated the
impact parameter dependence 
of the event-averaged electric flux that goes through the overlapping region, $ \Phi = \int_S
\vec E \cdot d \vec S $, which is shown in Fig.~\ref{fig:flux}. 
The plane $S$ is chosen so that it is perpendicular to the
line connecting the two centers of the two nuclei at $t = 0$, 
and it crosses the thickest part of the almond (dotted line in
Fig.~\ref{fig:coordinate}). 

In most central collisions, the electric flux is zero, and it gets 
larger when one increases $b$. 
For $1 \lesssim b \lesssim 5 \ [{\rm fm}]$, the flux is positive, which means that the field
directs toward the Au nucleus.  
At larger $b$ the electric flux changes its sign and the direction of
the fields is flipped. 
This result can be understood in the following way. 
At very peripheral collisions, the plane $S$ is closer to the center of
the Cu nucleus than to that of the Au nucleus.
As a result, the flux that comes from the Cu nucleus becomes denser,
because of its smaller radius. 
The behavior of the electric flux as a function of the impact parameter
would be reflected in the centrality dependence of charge-dependent $v_1$. 
}

Let us comment on
potential uncertainties in the estimate above.
It is possible that the charge dipole formed at the early
stage can be obscured in the later stages, namely the hydrodynamic evolution
and hadronic collisions.
In order to quantify these effects, 
we have to calculate the time evolution of the charge density
under an electric field.
The back reaction of matter to electromagnetic fields may also have to
be taken into account \cite{Tuchin:2010vs, Deng:2012pc}.
Hence, it would be desirable to use a magnetohydrodynamic model 
combined with a hadronic afterburner.
One should also consider the effects of fluctuations of the generated
electric fields on an event-by-event basis, although the fields have a
tendency to direct from Au to Cu nucleus on average.
The charge dipole could be weakened by the fluctuations.
Event-by-event simulations are necessary to consider the effect of
such fluctuations.
Finally, although we have assumed the dielectric constant is a constant,
it can in general depend on frequency and wave length.
Consideration of such effects is left as a future work.

\emph{Summary.}---
We have pointed out that, in Cu+Au collisions, 
a sizable strength of electric field directed from Au to Cu nucleus is generated in
the overlapping region.
We have shown this by performing event-by-event numerical
 calculation of the produced electromagnetic fields.
We have also pointed out that 
the electric field would induce an electric current in the matter created after
the collision and it would result in a dipole deformation of the charge
distribution in the medium.
We have shown that the 
charge-dependent directed flow of hadrons is 
sensitive to the charge dipole in the medium and 
is useful in estimating the electric conductivity of the QGP.

\acknowledgements
Y.~H. is grateful to A. Bzdak for helpful discussions.
The authors thank K.~Murase and Y.~Tachibana for useful discussions.
Y.~H. is supported by the Japan Society for the Promotion of Science for
Young Scientists. 
M.~H. is supported by RIKEN Junior Research Associate Program. 
The work of T.~H. is supported by
Grant-in-Aid for Scientific Research
No.~22740151.




\begin{thebibliography}{99}

\bibitem{Yagi:2005yb} 
  K.~Yagi, T.~Hatsuda and Y.~Miake,
  ``Quark-gluon plasma: From big bang to little bang,''
  Camb.\ Monogr.\ Part.\ Phys.\ Nucl.\ Phys.\ Cosmol.\  {\bf 23}, 1 (2005).

\bibitem{Adams:2005dq} 
  J.~Adams {\it et al.}  [STAR Collaboration],
  Nucl.\ Phys.\ A {\bf 757}, 102 (2005) [nucl-ex/0501009].

\bibitem{Adcox:2004mh} 
  K.~Adcox {\it et al.}  [PHENIX Collaboration],
  Nucl.\ Phys.\ A {\bf 757}, 184 (2005) [nucl-ex/0410003].

\bibitem{Romatschke:2007mq} 
  P.~Romatschke and U.~Romatschke,
  Phys.\ Rev.\ Lett.\  {\bf 99}, 172301 (2007) [arXiv:0706.1522 [nucl-th]].

\bibitem{Song:2010mg} 
  H.~Song, S.~A.~Bass, U.~Heinz, T.~Hirano, and C.~Shen,
  Phys.\ Rev.\ Lett.\  {\bf 106}, 192301 (2011)
  [Erratum-ibid.\  {\bf 109}, 139904 (2012)] [arXiv:1011.2783 [nucl-th]].

\bibitem{Gupta:2003zh} 
  S.~Gupta,
  Phys.\ Lett.\ B {\bf 597}, 57 (2004) [hep-lat/0301006].

\bibitem{Aarts:2007wj} 
  G.~Aarts, C.~Allton, J.~Foley, S.~Hands, and S.~Kim,
  Phys.\ Rev.\ Lett.\  {\bf 99}, 022002 (2007) [hep-lat/0703008 [HEP-LAT]].

\bibitem{Ding:2010ga} 
  H.~-T.~Ding, A.~Francis, O.~Kaczmarek, F.~Karsch, E.~Laermann, and W.~Soeldner,
  Phys.\ Rev.\ D {\bf 83}, 034504 (2011). [arXiv:1012.4963 [hep-lat]].

\bibitem{Buividovich:2010tn} 
  P.~V.~Buividovich, M.~N.~Chernodub, D.~E.~Kharzeev, T.~Kalaydzhyan, E.~V.~Luschevskaya and M.~I.~Polikarpov,
  Phys.\ Rev.\ Lett.\  {\bf 105}, 132001 (2010) [arXiv:1003.2180 [hep-lat]].

\bibitem{Arnold:2003zc} 
  P.~B.~Arnold, G.~D.~Moore, and L.~G.~Yaffe,
  JHEP {\bf 0305}, 051 (2003) [hep-ph/0302165].


\bibitem{Huang:2012sc} 
  S.~Huang [PHENIX Collaboration],
  Nucl.\ Phys.\ A {\bf 904-905}, 417c (2013)
  [arXiv:1210.5570 [nucl-ex]].

\bibitem{Bzdak:2011yy} 
  A.~Bzdak and V.~Skokov,
  Phys.\ Lett.\ B {\bf 710}, 171 (2012) [arXiv:1111.1949 [hep-ph]].

\bibitem{Kubo:1957mj} 
  R.~Kubo,
  J.\ Phys.\ Soc.\ Jap.\  {\bf 12}, 570 (1957).

\bibitem{McLerran:1984ay}
  L.~D.~McLerran and T.~Toimela,
  Phys.\ Rev.\  D {\bf 31}, 545 (1985).

\bibitem{Gale:1987ki} 
  C.~Gale and J.~I.~Kapusta,
  Phys.\ Rev.\ C {\bf 35}, 2107 (1987).

\bibitem{Forster:1990} 
D.~Forster, Hydrodynamics, Fluctuations, Broken Symmetry, and Correlation Functions (Perseus Books, New York, 1990).

\bibitem{Akamatsu:2011nr} 
  Y.~Akamatsu, H.~Hamagaki, T.~Hatsuda, and T.~Hirano,
  Phys.\ Rev.\ C {\bf 85}, 054903 (2012)
  [arXiv:1107.3612 [nucl-th]].

\bibitem{Fukushima:2008xe} 
  K.~Fukushima, D.~E.~Kharzeev, and H.~J.~Warringa,
  Phys.\ Rev.\ D {\bf 78}, 074033 (2008)
  [arXiv:0808.3382 [hep-ph]].

\bibitem{Burnier:2011bf} 
  Y.~Burnier, D.~E.~Kharzeev, J.~Liao, and H.~-U.~Yee,
  Phys.\ Rev.\ Lett.\  {\bf 107}, 052303 (2011) [arXiv:1103.1307 [hep-ph]].

\bibitem{Burnier:2012ae} 
  Y.~Burnier, D.~E.~Kharzeev, J.~Liao, and H.~-U.~Yee,
  arXiv:1208.2537 [hep-ph].

\bibitem{Deng:2012pc} 
  W.~-T.~Deng and X.~-G.~Huang,
  Phys.\ Rev.\ C {\bf 85}, 044907 (2012) [arXiv:1201.5108 [nucl-th]].

\bibitem{Alver:2008aq} 
  B.~Alver, M.~Baker, C.~Loizides, and P.~Steinberg,
  arXiv:0805.4411 [nucl-ex].

\bibitem{Schwinger:1951nm} 
  J.~S.~Schwinger,
  Phys.\ Rev.\  {\bf 82}, 664 (1951).

\bibitem{Tuchin:2010vs} 
  K.~Tuchin,
  Phys.\ Rev.\ C {\bf 82}, 034904 (2010)
  [Erratum-ibid.\ C {\bf 83}, 039903 (2011)]
  [arXiv:1006.3051 [nucl-th]].

\end{thebibliography}
\end{document}